%Paper: hep-th/9510165
%From: "Oleg K. Sheinman" <sheinman@landau.ac.ru>
%Date: Sat, 21 Oct 1995 21:20:54 +0300
%Date (revised): Thu, 9 Nov 1995 19:47:06 +0300

\input amstex
\documentstyle{amsppt}
\topmatter

\title INTEGRABLE MANY-BODY SYSTEMS OF CALOGERO-MOSER-SUTHERLAND TYPE
                    IN HIGH DIMENSION
\endtitle

\author               O.K.Sheinman
\endauthor

\thanks
   To appear in International Mathematical Research
                Notices volume 1995
\endthanks

\NoRunningHeads
\endtopmatter
\NoBlackBoxes
\TagsOnRight
\define \fh{{\frak h}}
\define \fg{{\frak g}}
   % Krichever-Novikov algebra
\define \CalG{{\Cal G}}   % Krichever-Novikov algebra
\document

The many-body problems of Calogero-Moser-Sutherland  type have a long
history
now. They are used in different areas of physics including conformal
field theory and gauge theories. The detail review of the problem and
references can be found in [1]. In recent papers [1,2] a treatment of
elliptic Calogero-Moser-Sutherland  system was suggested, which allows
to obtain this
system together with the complete set of its integrals starting up
from the interacting Higgs field and vector field on the elliptic
curve (the complete integrability of the system was obtained in [3]
using inverse scattering technique). The method of [1,2] is the one
suggested in [4,5], i.e. the method of Hamiltonian reduction.
{}From the more general point of view [2] such type of constructions
is connected with some moduli space of holomorphic $G$-bundles over
smooth Riemann surfaces, where $G$ is semisimple finite-dimensional
Lie group.

We suggest in this paper a construction which is very close to the
one of [1,2] but which is connected with Krichever-Novikov (KN)
algebras of affine type which were introduced in [6] in connection
with quantization of multiloop string diagrams and soliton theory.

We consider a system, which consists of a vector field and of a Higgs
field with a free hamiltonian on a Riemann surface of nonzero genus.
In mathematical words we consider an element of a KN-algebra and an
element of its current subalgebra.  Being subjected to the
Hamiltonian reduction procedure that system provides a kind of
mani-body problem which turns out to be the problem of
Calogero-Moser-Sutherland type. The dimension of its configuration space
equals to the genus of the Riemann surface under consideration.  One
of the fields under consideration provides the configuration
parameters of a certain many-body problem, while the invariants of
the other field provide its integrals.  It turns out (Theorem 2.1
below) that the number of configuration parameters is equal to the
number of invariants, which proves the complete integrability of the
obtained many-body problem in Liouville sense.  In case of an
elliptic curve we come to the result of [1,2] on the
Calogero-Moser-Sutherland systems in elliptic case.

The author is grateful to I.Krichever for the fruitful discussions.

\head        1. Krichever-Novikov algebras of affine type
\endhead
Let $\Gamma$ be a compact algebraic curve over ${\Bbb C}$ with two
marked points $P_{\pm},$ ${\Cal A}^\Gamma$ be the algebra of
meromorphic functions on $\Gamma$ regular outside the points $P_\pm ,$
$e$ be a vector field on $\Gamma$ with the same analitical
properties, ${\frak g}$ be a complex semisimple Lie algebra,
    $${\Cal G}={\frak g}\otimes _{\Bbb C}{\Cal A}^{\Gamma }
       \oplus {\Bbb C}c\oplus {\Bbb C}\nabla               \tag 1.1
    $$
be a KN-algebra of affine type [6,7] with the
one-dimensional centre generated by $c,$
$\nabla$ be the dirivative along the vector field $e.$ The
elements of the algebra ${\Cal G}$ will be denoted by ${\tilde
X}=X+ac+be,$ where $X\in{\frak g}\otimes _{\Bbb C}{\Cal A}^\Gamma ,$
$a,b\in{\Bbb C}.$
${\Cal G}$ is considered to be identified with its dual
space ${\Cal G}^*$ by means the nondegenerate form
      $$<X+ac+be,\,Y+a^\prime c+b^\prime e>={1\over{2\pi i}}
         \oint\limits _{c_0}(X,Y){{\text{d}p}\over{E(p)}}+
         ab^\prime +ba^\prime ,                           \tag 1.2
      $$
where $(\cdot ,\cdot)$ is the Cartan-Killing form on $\frak g$ and
in local coordinates $e=E(p)({\text{d}p})^{-1}$.
A structure of the central extension on ${\Cal G}$ is given by means
the cocycle
       $$ \gamma (X,Y)={1\over{2\pi i}}
           \oint\limits _{c_0}(X,\text{d}Y).              \tag 1.3
       $$
It should be mentioned that the analog of the relation (1.2) in [1]
(see relation (4.1) there) necessarily has an arbitrariness in the
choice of an integration 2-form, so it can not be generalized on the
case $g>1$ in an invariant way.

{\it Adjoint action:}
     $$ \split
        (\text{Ad}g)(X+ac+be)=ac+be+gXg^{-1}-b(\nabla g)g^{-1}\\
        \,\,\,\, +(<g^{-1}(\nabla g),X>-{b\over 2}
        <(\nabla g)g^{-1},(\nabla g)g^{-1}>)c.  \endsplit
                                                          \tag 1.4
     $$
In order to classify the {\it orbits} of adjoint action let us
consider the monodromy equation
   $$(\nabla +X)\Psi =0                                   \tag 1.5
   $$
and let $\Psi _X$ be the multivalued solution of this equation
satisfying the initial condition $\Psi _X(\gamma _0)=1$ in some fixed
point $\gamma _0\in\Gamma $. Let $G_X$ be the corresponding monodromy
group.
\proclaim{Theorem 1.1 [7]} Two elements $\nabla +X,\nabla +Y\in
 \CalG$ belong to the same orbit of adjoint action if and only if
 $<X,X>=<Y,Y>$ and
 there exists such a $g\in G$ that $gG_Xg^{-1}=G_Y$ (where
 $G=\exp\fg$).
\endproclaim

If $G_X$ is an abelian group, the corresponding orbit is said to be
{\it commutative}. In that case up to conjugation one has
   $$\Psi _X=\exp\int\limits _{\gamma _0}^\gamma\omega _X,
     \,\,\text{where}\,\,
     \omega _X=-(\text{d}\Psi _X)\Psi _X^{-1} \in\fh\otimes\Omega _1
     (\Gamma )                                              \tag 1.6
   $$
and $\Omega _1(\Gamma )$ is the space of 1-forms on $\Gamma$ which
are meromorphic on $\Gamma$ and regular out the points $P_\pm .$ The
relation (1.6) establishes the {\it duality} between the currents $X$
and differentials $\omega _X.$ In other way the same duality can be
established by means the nondegenerate form (1.2).
\proclaim{Lemma 1.1} The duality between currents and differentials
 established by means the belinear form (1.2) coincides with the
 duality established by the relation (1.6) and in local coordinates
 looks as follows:
   $$X{{\text{d}p}\over{E(p)}}=\omega _X.                   \tag 1.7
   $$
\endproclaim

The {\it Weil group} $W$ is defined as the group classifying
commutative orbits.  Let $Q^\vee$ be the dual root lattice and
${\overline W}$ be the Weil group of Lie algebra $\fg ,$ then
$W={\overline W}(Q^\vee)^{2g+1}$ [7]. The group $W$ acts on the space
$\fh\otimes\Omega _1(\Gamma)$ by means transformating the periods of
differentials of that space.

{}From now on we shall suppose that $\Omega _1$ is the
$2g+1$-dimensional linear space which consists of such 1-forms on
$\Gamma$ that are holomorfic outside the points $P_\pm ,$ and have
the orders not less then $-1$ and $-g$ in the points $P_+$ and $P_-$
respectively.

\proclaim{Theorem 1.2 [7]} The space of commutative orbits is
isomorphic to the quotient $\Omega =\fh\otimes\Omega _1/W.$
\endproclaim

\head      2. The dynamical system
\endhead
A point of the {\it phase space} consists of the element  $\nabla
+X\in\CalG$ ("vector field" in physical terms), of a function $\phi
:\Gamma\rightarrow\fg$ with an unique exponential singularity in
$P_-,$ and of a $\fg$-valued 1-form $\omega $ of the
$(2g+1)\times\dim\fg$-dimensional space $\fg\otimes\Omega _1.$
We shall suppose $X$ to satisfy the relation
$\text{res}_{P_+}(X{{\text{d}p}\over{E(p)}})=0.$

The {\it gauge action is as follows}:
   $$X\rightarrow gXg^{-1}-(\nabla g)g^{-1},\,\,\,
     \phi\rightarrow g\phi g^{-1},\,\,\,
     \omega\rightarrow g\omega g^{-1}-(\text{d}g)g^{-1},   \tag 2.1
   $$
where $g$ is a group current.

The {\it symplectic form}. Let $X\mapsto\omega _X=X{{\text{d}p}\over
E},$ $A_k^X$ and $B_k^X$ be the a-periods and the b-periods of
$\omega _X$ respectively, $A_k^\omega$ and $B_k^\omega$ be the same
for the 1-form $\omega$ $(k=1,\ldots ,g).$ The symplectic form $S$ is
defined by the relation
   $$S={1\over{2\pi i}}\oint\limits_{c_0}(\delta X\wedge\delta\phi)
     {{\text{d}p}\over E} +{1\over 2}\sum\limits _{j=1}^g
     (\delta A_j^X\wedge\delta B_j^\omega +
      \delta A_j^\omega \wedge\delta B_j^X),                \tag 2.2
   $$
where for any $\delta X,\,\delta Y\in\fg$ one has
  $$\delta X\wedge\delta Y=
      \sum\limits _\alpha \delta X_\alpha\wedge\delta Y_{-\alpha}
     +\sum\limits _{k=1}^l \delta X_{\alpha_k}\wedge\delta
     Y_{\alpha_k},                                          \tag 2.3
  $$
and
   $$\delta X =\sum\limits_\alpha \delta X_\alpha e_\alpha +
     \sum\limits _{k=1}^l \delta X_{\alpha_k}h_{\alpha_k},
   $$
   $$\delta Y =\sum\limits_\alpha \delta Y_\alpha e_\alpha +
     \sum\limits _{k=1}^l \delta Y_{\alpha_k}h_{\alpha_k}
   $$
are the canonical decompositions of the elements $\delta X,\,\delta Y$
respectively ($l=rank\,\fg ,$ $\alpha$ being the roots of $\fg ,$
$\alpha _k$ being the simple roots of $\fg $ and $h_{\alpha _k}$
being their dual elements ($k=1,\ldots ,l$)).

The {\it moment map,} which corresponds to (2.2), reads
   $$\mu (X,\phi ,\omega )=[\nabla +X,\phi ]+
      {1\over 2}\sum\limits _{j=1}^g [A_j^X,B_j^\omega ]+
      {1\over 2}\sum\limits _{j=1}^g [A_j^\omega ,B_j^X].     \tag 2.4
   $$

We put the {\it hamiltonian} of the system to be equal
   $$H=\sum\limits _{j=1}^g (A_j^\omega ,A_j^\omega )
      +g\sum\limits _{\alpha\in R_+}\phi _\alpha \phi _{-\alpha},
                                                              \tag 2.5
   $$
where $R$ is the root system of $\fg$ and $\phi = \sum\limits
_{\alpha\in R}\phi _\alpha e_\alpha +\sum\limits _{k=1}^l \phi
_kh_{\alpha _k}$ is the canonical decomposition of the element
$\phi\in\fg .$

Now we are in position to carry out a kind of {\it hamiltonian
reduction} which results in the required many-body problem. First of
all let us restrict our phase space by the two $g$-invariant
conditions:
 \roster
  \item"{$1^0$}" $\omega _X=\omega$
  \item"{$2^0$}" monodromy group $G_X$ is abelian.
 \endroster
\remark{Remark 2.1} Denote $\Psi =\exp\int\limits _{\gamma _0}^\gamma
 \omega ,$ then the conditions $1^0,$ $2^0$ imply
   $$(\nabla +X)\Psi =0.                                   \tag 2.6
   $$
\endremark
Now we want to reduce the system on the zero level of the moment map,
i.e. on the manifold $\mu (X,\phi ,\omega )=0.$ Because of the $2^0$
both sums in (2.4) vanish and {\it the equation of zero moment} reads
   $$[\nabla +X,\phi]=0                                  \tag 2.7
   $$
\remark{Remark 2.2} Both equations (2.6) and (2.7) appear in the
Krichever's algebraic-geometrical construction of the solutions of
zero curvature equations [10].
\endremark

The equation (2.7) can be rewritten in the form
   $$\nabla\phi +(\text{ad}X)\phi =0.                     \tag 2.8
   $$
For $\phi =\sum\limits _{\alpha\in R}\phi _\alpha e_\alpha +
\sum\limits _{k=1}^l \phi _kh_{\alpha _k}$  one has $(\text{ad}X)\phi
=\sum\limits _{\alpha\in R}\alpha (X)\phi _\alpha .$ Furtheremore
$\alpha (X)$ satisfies the relation $\alpha (X{{\text{d}p}\over E})
=\omega _X (h_\alpha,)$ $h_\alpha\in\fh$ ($h_\alpha $ is uniquely
defined). So the vector equation (2.7) (which appears as the matrix
equation in case of classical Lie algebras $\fg$) is equal to the
following system of scalar equations:
   $$ \text{d}\phi _\alpha +\omega _X(h_\alpha)\phi _\alpha =0
      \,\,\,(\alpha\in R,\,\alpha\ne 0 ),                \tag 2.9
   $$
   $$ \text{d}\phi _\alpha =0\,\,\,
       (\alpha\in R,\,\alpha =0 ).                       \tag 2.10
   $$
The 1-form $\omega _X(h_\alpha )$ is ${\Bbb C}$-valued, so the
equation (2.9)has the unique up to the scalar factor solution, namely
the Baker-Achieser function
     $$ \phi _\alpha (\gamma)=(\exp\int\limits _{\gamma _0}^\gamma
        \omega _X (h_\alpha)) {{\theta (A(\gamma)+Z(D)+U_\alpha)}\over
        {\theta (A(\gamma)+Z(D))\theta (Z(D)+U_\alpha )}},
                                                          \tag 2.11
     $$
where $\gamma ,\gamma _0\in\Gamma ,$ and $\gamma _0\ne P_\pm$  is
fixed, $\theta $ is the Riemann theta-function, $A$ is the
transformation of Abel,
$Z(D)=-A(D)-{\Cal K}$ (${\Cal K}$ being the the vector
of Riemann constants), and $U_\alpha$  is the vector of b-periods of
the differential $\omega _X(h_\alpha ).$ Note that the factor $\theta
((Z(D)+U_\alpha ))$ in the denominator of (2.11) is simply a
constant. The reason of the choice of this constant will be clear
at once.
\remark{Remark 2.3} It will be shown in Section 3 that the factor
$\theta (Z(D)+U_\alpha )$ is also related to the Weil-invariance of
the hamiltonian.
\endremark
\proclaim{Lemma 2.1} If $\gamma =\gamma _0$ and $\alpha\ne 0$ then
$\phi _\alpha
(\gamma _0)={1\over{\theta (Z(D))}},$ and $\sum\limits _{\alpha\in
R}\phi _\alpha (\gamma _0)e_\alpha = {1\over{\theta (Z(D))}}J,$
where $J=\sum\limits _{\beta\in R_+}(e_\beta + e_{-\beta}).$
\endproclaim
\demo{Proof} If $\gamma =\gamma _0$ then $\exp\int\limits _{\gamma
_0}^{\gamma} \omega _X(h_\alpha )=1,$ $A(\gamma )=0$ and so
$\phi _\alpha (\gamma _0)={{\theta (Z(D)+U_\alpha )}\over{\theta
(Z(D))\theta (Z(D)+U_\alpha )}}={1\over{\theta (Z(D))}}$ independently
of $\alpha .$
\enddemo
\example{Example 2.1} Let $\fg =\fg{\frak l}(N).$
 Then
   $$J=\left(\matrix
          0   &    1   & \hdots &    1    \\
          1   &    0   & \hdots &    1    \\
       \vdots & \vdots & \ddots & \vdots  \\
          1   &    1   & \hdots &    0
       \endmatrix\right).
   $$
\endexample
It remains only to resolve the equation (2.10) in order to complete
the reduction. One has from (2.10) $\phi _\fh =const$ $(k=1,\ldots
,l),$ where $\phi _\fh =\sum\limits _k\phi _kh_{\alpha _k}.$ Let us
correspond to $j$-th handle of $\Gamma$ its own constant $\phi _\fh
^{(j)}$ and respectively its own solution  $\phi ^{(j)}$ of
(2.9)-(2.10), which satisfy the condition
   $${1\over{\theta (Z(D))}}J +A_j^\omega =\phi ^{(j)}(\gamma _0).
                                                             \tag 2.12
   $$
in any gauge where $X\in\fh .$ This relation does not depend on the
choice of such kind a gauge. In fact $A_j^\omega$ and $\phi
^{(j)}(\gamma _0)$ are transformed by the low $(\cdot )\mapsto
g_0(\cdot )g_0^{-1},$ where $g_0=g(\gamma _0).$ If
$(\text{Ad}g_0)\fh =\fh$ then the action of $g_0$ reduces to the
action of an element of the Weil group of $\fg ,$ but $J$ is
Weil-invariant.

As in the gauge where $X\in\fh$ noncartanian part
of $\phi (\gamma _0)$ is $constant$ and is equal to $J/\theta (Z(D))$
(Lemma 2.1), all the conditions (2.12) are compatible and have the
solution
   $$  A_j^\omega =\phi _\fh ^{(j)}\,\,(j=1,\ldots ,g).       \tag 2.13
   $$
\proclaim{Lemma 2.2} The integral term on the right hand in (2.2)
vanishes on the reduced phase space.
\endproclaim
\demo{Proof} In the gauge where $\delta X\in\fh$ one has
$\delta X\wedge\delta\phi =\delta X\wedge\delta\phi _\fh$ (see
(2.3)). As it was just now shown, $\phi _\fh$ is a constant in
$\gamma .$ Furthermore
   $${1\over{2\pi i}}\oint\limits_{c_0}(\delta X\wedge\delta\phi
     _\fh) {{\text{d}p}\over E} =
     {1\over{2\pi i}}(\oint\limits_{c_0}\delta X{{\text{d}p}\over E})
     \wedge\delta\phi _\fh =
     \delta ({1\over{2\pi i}}\oint\limits_{c_0}X{{\text{d}p}\over E})
     \wedge\delta\phi _\fh .
   $$
There is nothing but
$\delta (\text{res}_{P_+}X{{\text{d}p}\over E})\wedge\delta\phi_\fh$
on the right hand of the last relation.
But $\text{res}_{P_+}X{{\text{d}p}\over E}=0,$
therefore its variation vanishes.
\enddemo

In order to summarize the obtained results let us introduce a new
notation. Denote
   $$P_j=A_j^X=A_j^\omega ,\,\, Q_j=B_j^X=B_j^\omega
     \,(j=1,\ldots ,g)                                     \tag 2.14
   $$
and define the vectors $U_k\in{\Bbb C}^g$ $(k=1,\ldots ,l)$ from the
relation
   $$(Q_1,\ldots ,Q_g)=(U_1,\ldots ,U_l)^T.                 \tag 2.15
   $$
$U_k$ will be considered as the vector of the b-periods of the
differential $\omega (h_{\alpha _k})$ $(k=1,\ldots ,l),$ where
$\alpha _1,\ldots ,\alpha _l$ are the simple roots of $\fg .$ For an
arbitrary root $\alpha$ the vector $U_\alpha$ and $U_1,\ldots ,U_l$
are related as follows: if $\alpha =\sum\limits _{k=1}^lm_k\alpha _k$
($m_k$ are nonnegative integers), then
   $$U_\alpha =\sum\limits _{k=1}^lm_kU_k.                 \tag 2.16
   $$

The procedure of hamiltonian reduction, which has been carried out in
this Section, results in the following dynamical system, which will
be called {\it the reduced system}. The phase parameters of the
reduced system are as follows: $P_j,Q_j\in\fh$  $(j=1,\ldots ,g).$
The symplectic structure is given by the standard relation
   $$S_{red}=\sum\limits _{j=1}^g P_j\wedge Q_j           \tag 2.17
   $$
(it follows from (2.2) because of (2.14) and Lemma 2.2).
Finally the hamiltonian of the reduced system is of the form
   $$H=\sum\limits _{j=1}^g \sum\limits _{k=1}^l \phi _{jk}^2 +
     g\sum\limits _{\alpha >0}\phi _\alpha\phi _{-\alpha}  , \tag 2.18
   $$
where $\phi _\alpha$ is given by (2.11), $U_\alpha$ are related to
the phase parameters by (2.15) and (2.16) and $\phi _{jk}$ are
defined by the relation $\phi _\fh ^{(j)}=\sum\limits _{k=1}^l \phi
_{jk}h_{\alpha _k}$ $(j=1,\ldots ,g).$
\example{Example 2.2} Let $\fg =\fg{\frak l}(N).$ Then $l=N,$ an
 arbitrary root is of the form $\alpha =\alpha _{ik}(h)=h_i-h_k$
 $(h=diag(h_1,\ldots ,h_N)),$ $\omega (h_{\alpha _{ik}})=\omega
 _i-\omega _k$ $(\omega =diag(\omega _1,\ldots ,\omega _N))$ and
 therefore $U_\alpha =U_i-U_k.$ Therefore (2.18) reads
   $$H=\sum\limits _{j=1}^g (P_j,P_j) +
       g\sum\limits _{i<k}{
        {\theta (A(\gamma)+Z(D)+U_i-U_k)\theta
        (A(\gamma)+Z(D)+U_k-U_i)}
        \over
        {\theta (A(\gamma)+Z(D))^2\theta (Z(D)+U_i-U_k)
        \theta (Z(D)+U_k-U_i)}},                           \tag 2.19
   $$
where $U_i$ $(i=1,\ldots ,N)$ are related with $Q_j$
($j=1,\ldots ,g$)
via (2.15),(2.16). It is evident, that (2.19) is the many-body problem
in $g$-dimensional space with the pairwise interaction.
\endexample
\proclaim{Theorem 2.1} The reduced hamiltonian system with the
hamiltonian (2.18) and the symplectic structure (2.17) is completely
integrable in sense of the Liouvill theorem.
\endproclaim
\demo{Proof} By definition $P_j,Q_j\in \fh$ $(j=1,\ldots ,g)$ and
$\dim\fh =l.$ Therefore the dimension of the phase space is equal to
$2gl.$

On the other hand for each $j=1,\ldots ,g$ the Chevalley invariants
of the $\fg$-valued function $\phi ^{(j)}$ appear as the integrals of
the system. For each $j$ the function $\phi ^{(j)}$ has $rank\,\fg
=l$ invariants, so the system has $gl$ independent integrals.
Applying instead the basic symmetric functions of $g$ variables to
each set of the same Chevalley invariants differing by $j$ only one
can obtain the family of $gl$ independent integrals in involution.
The Theorem is proved.
\enddemo

In conclusion let us consider the example 2.2 with $g=1$ in more
detail.
\example{Example 2.3} Let $\fg =\fg{\frak l}(N),$ $g=1$ ($\Gamma$ is
an elliptic curve). Then by Abel transformation $\Gamma$ can be
identified with the torus ${\Bbb T}^2.$ In other words one can denote
$A(\Gamma )=z\in{\Bbb T}^2.$ As $g=1$ then by choice of the initial
point $\gamma _0$ the constant $Z(D)$ can be vanished. Thus in case
$g=1$
   $$H=(P_1,P_1) +
        \sum\limits _{i<k}{
        {\theta (z+U_i-U_k)\theta (z-U_i+U_k)}
        \over
        {\theta (z)^2\theta (U_i-U_k)^2}}.                  \tag 2.20
   $$
Because of the addition theorem
   $$ {{\theta (z+u)\theta (z-u)}
        \over
      {\theta (z)^2\theta (u)^2}}={\frak P}(u)-{\frak P}(z),
   $$
where ${\frak P}$ is the Weierstrass ${\frak P}$-function. Therefore
   $$H=(P_1,P_1) +
        \sum\limits _{i<k} ({\frak P}(U_i-U_k)-{\frak P}(z)),
                                                            \tag 2.21
   $$
that coincides with {\it elliptic Calogero--Moser--Sutherland system}
(cf. [1,2]).
\endexample
\remark{Remark 2.4}The reduced symplectic structure (2.17) is
exactly a kind of the symplectic structure due to Howe, which is
entensively exploated in the number theory investigations [8].
\endremark

\head          3. Weil-invariance of the reduced hamiltonian
\endhead
The phase space of the reduced dynamical system (2.17)-(2.18) is
endowed with the natural $W$-action, where $W$ is the Weil group of
${\Cal G}$ (see Section 1). Now we want to return to the Remark 2.3
and to show that the choice of the scalar factor
$\theta (Z(D)+U_\alpha )$ in (2.11) is related to the
Weil-invariance of the hamiltonian.

\proclaim{Theorem 3.1} The hamiltonian $H,$ defined by (2.18), is
 $W$-invariant.
\endproclaim

\demo{Proof} Let ${\overline W}$ be the Weil group of the Lie algebra
 $\fg ,$ $L$ be the lattice of periods of the Riemann surface $\Gamma
 .$ As it was above mentioned (Sect. 1) the action of $W$ on
 the space $\fh\otimes\Omega _1(\Gamma)$ results, first, in the
 translations of periods of 1-forms by elements of $L,$ and, second,
 in the action of ${\overline W}$ on $\fh$-cofactor. Each of the
 summands in (2.18) is evidently ${\overline W}$-invariant, so it
 suffices to prove its $L$-invariance.

 Let $\omega\in\fh\otimes\Omega _1(\Gamma).$ Translating the periods
 of $\omega$ by the elements of $L$ is equivalent to the shift
 $\omega (h_\alpha )\mapsto \omega (h_\alpha )+\omega _{int},$ where
 $\omega _{int}=\sum\limits _{k=1}^g n_k\omega _k,$ $\omega _k$ are
 the basic holomorphic differentials $(k=1,\ldots ,g).$ It remains to
 prove the invariance of $\phi _\alpha$ in respect to such kind a
 shift and, moreover, it suffices to take $\omega _{int}=\omega _k$
 $(k=1,\ldots ,g).$

 Note that $\exp\int\limits _{\gamma _0}^\gamma\omega _k=\exp (2\pi
 iA_k(\gamma ))$ ($A_k$ is the $k$-th component of the Abel
 transformation). Moreover, under the above mentioned shift
 $U_\alpha\mapsto U_\alpha +{1\over{2\pi i}}\oint\limits _b\omega _k
 =U_\alpha +B_k$ (where $B_k$ is the $k$-th column of the matrix of
 b-periods). The latest mapping results in the multiplying the
 $\theta (A(\gamma )+Z(D)+U_\alpha)$ by $\exp(-\pi i(B_{kk}+
 2A_k(\gamma )+Z_k(D)+({\overline U}_\alpha )_k))$ in (2.11) and in
 the multiplying the $\theta (Z(D)+U_\alpha )$ by $\exp (-\pi
 i(B_{kk}+2(U_\alpha )_k+Z_k(D)))$ (index $k$ denotes the $k$-th
 coordinate of a vector). Thus the ratio of $\theta$'s in (2.11) gets
 the factor $\exp (-2\pi i A_k(\gamma )),$ that compensates the
 variation of the $\exp\int\limits _{\gamma _0}^\gamma\omega
 (h_\alpha ).$ Thus $\phi _\alpha$ is $L$-invariant for each $\alpha
 ,$ and because for (2.18) $H$ is gauge invariant.
\enddemo

%\newpage
\end